# GPGPU COMPUTING

**BOGDAN OANCEA**[*]
**TUDOREL ANDREI**[**]
**RALUCA MARIANA DRAGOESCU**[***]

**Abstract**
*Since the first idea of using GPU to general purpose computing, things have evolved over the years and now there are several approaches to GPU programming. GPU computing practically began with the introduction of CUDA (Compute Unified Device Architecture) by NVIDIA and Stream by AMD. These are APIs designed by the GPU vendors to be used together with the hardware that they provide. A new emerging standard, OpenCL (Open Computing Language) tries to unify different GPU general computing API implementations and provides a framework for writing programs executed across heterogeneous platforms consisting of both CPUs and GPUs. OpenCL provides parallel computing using task-based and data-based parallelism. In this paper we will focus on the CUDA parallel computing architecture and programming model introduced by NVIDIA. We will present the benefits of the CUDA programming model. We will also compare the two main approaches, CUDA and AMD APP (STREAM) and the new framwork, OpenCL that tries to unify the GPGPU computing models.*

**Keywords**: *CUDA, Stream, OpenCL, GPU computing, parallel computing*

**Introduction**

Parallel computing offer a great advantage in terms of performance for very large applications in different areas like engineering, physics, biology, chemistry, computer vision, econometrics. Since the first supercomputers in early '70s, the nature of parallel computing has changed and new opportunities and challenges have appeared over the time. While 30-35 years ago computer scientists used massively parallel processors like the Goodyear MPP[1] Connection Machine[2], Ultracomputer[3] and machines using Transputers[4] or dedicated parallel vector computers, like Cray computer series, nowadays off-the-shelves desktops have FLOP rates greater than a supercomputer in late 80's. For example, if Cray 1 had a peak performance of 80 Megaflops and CRAY X MP had a peak performance of 200 MFLOPS an actual multicore Intel processor has more than 120 GFLOPS. These figures emphasize a major shift in computer and processors design.

The processors speed increased as the clock frequency and the number of transistors increased over the time. The clock frequency has increased by almost four orders of magnitudes between the first 8088 Intel processor and the actual processors, and the number of transistors also raised from 29.000 for Intel 8086 to approximately 730 million for an Intel Core i7-920 processor[5]. The increased clock frequency has an important side effect - the heat dissipated by processors. To overcome this problem, instead of increasing the clock frequency the processor designers come with a new

---

[*] Professor, Ph. D., "Nicolae Titulescu" University (email: bogdan.oancea@gmail.com).
[**] Professor, Ph. D., Bucharest Academy of Economic Studies (email: andreitudorel@yahoo.com).
[***] Assistant Lecturer, "Artifex" University (email: raluca_dragoescu@yahoo.com).
[1] K. E. Batcher, (1980), Design of a Massively Parallel Processor, *IEEE Transactions on Computers*, Vol. C29, September, pp. 836–840.
[2] Lewis W. Tucker, George G. Robertson, (1988), Architecture and Applications of the Connection Machine, *Computer*, vol. 21, no. 8, pp. 26–38.
[3] Allan Gottlieb, Ralph Grishman, Clyde P. Kruskal, Kevin P. McAuliffe, Larry Rudolph, Marc Snir, (1982), *The NYU Ultracomputer—designing a MIMD, shared-memory parallel machine*, ISCA '82 Proceedings of the 9th annual symposium on Computer Architecture, pp. 27 – 42.
[4] Barron, Iann M. (1978), D. Aspinall. ed. "The Transputer". The Microprocessor and its Application: an Advanced Course, Cambridge University Press.
[5] INTEL (2012), Microprocessor Quick reference guide.

paradigm – multicore processors. Both INTEL and AMD offer multicore processors that are now common for desktop computers. Multicore processors turn normal desktops in truly parallel computers. Although the computing power of multicore processors is amazing, new applications demand more and more computational power.

In 2003, Mark Harris[6] recognized the potential of using graphical processing units (GPU) for general purpose applications. Modern GPUs are high performance many-core processors that can obtain high FLOP rates. In the past the processing units of the GPU were designed only for computer graphics but now GPUs are truly general-purpose parallel processors. Since the first idea of using GPU for general purpose computing, GPU programming models have evolved and there are several approaches to GPU programming now: CUDA (Compute Unified Device Architecture) from NVIDIA and APP (Stream) from AMD. A great number of applications were ported to use the GPU and they obtain speedups of few orders of magnitude comparing to optimized multicore CPU implementations.

GPGPU (general-purpose computing on graphics processing units) is used nowadays to speed up parts of applications that require intensive numerical computations. Traditionally, these parts of applications are handled by the CPUs but GPUs have now MFLOPs rates much better than CPUs[7]. The reason why GPUs have floating point operations rates much better even than multicore CPUs is that the GPUs are specialized for highly parallel intensive computations and they are designed with much more transistors allocated to data processing rather than flow control or data caching[8].

Khronos Group's OpenCL (Open Computing Language)[9] is a new emerging standard, that tries to unify different GPU general computing API implementations like CUDA or APP (Stream) and to provide a general framework for writing programs executed across heterogeneous platforms consisting of both CPUs and GPUs.

**The CUDA programming model**

Figures 1 and 2[10] shows the advances of the current GPUs that become highly parallel, multithreaded, manycore processors with an amazing computational power and very high memory bandwidth. Figure 1 plots the FLOP rates of GeForce GPUs compared with Intel processors while figure 2 plots the memory bandwidth of the NVIDIA GPUs.

---

[6] Harris, Mark J., William V. Baxter III, Thorsten Scheuermann, and Anselmo Lastra.( 2003), Simulation of Cloud Dynamics on Graphics Hardware. In *Proceedings of the IGGRAPH/Eurographics Workshop on Graphics Hardware* 2003, pp. 92-101.
[7] NVIDIA, (2011), NVIDIA CUDA C Programming Guide, version 4.0.
[8] NVIDIA, (2010), CUDA C Programming Guide, Version 4.0
[9] Khronos OpenCL Working Group (2009), The OpenCL Specification - Version 1.0. The Khronos Group, Tech. Rep
[10] NVIDIA, (2011), NVIDIA CUDA C Programming Guide, version 4.0

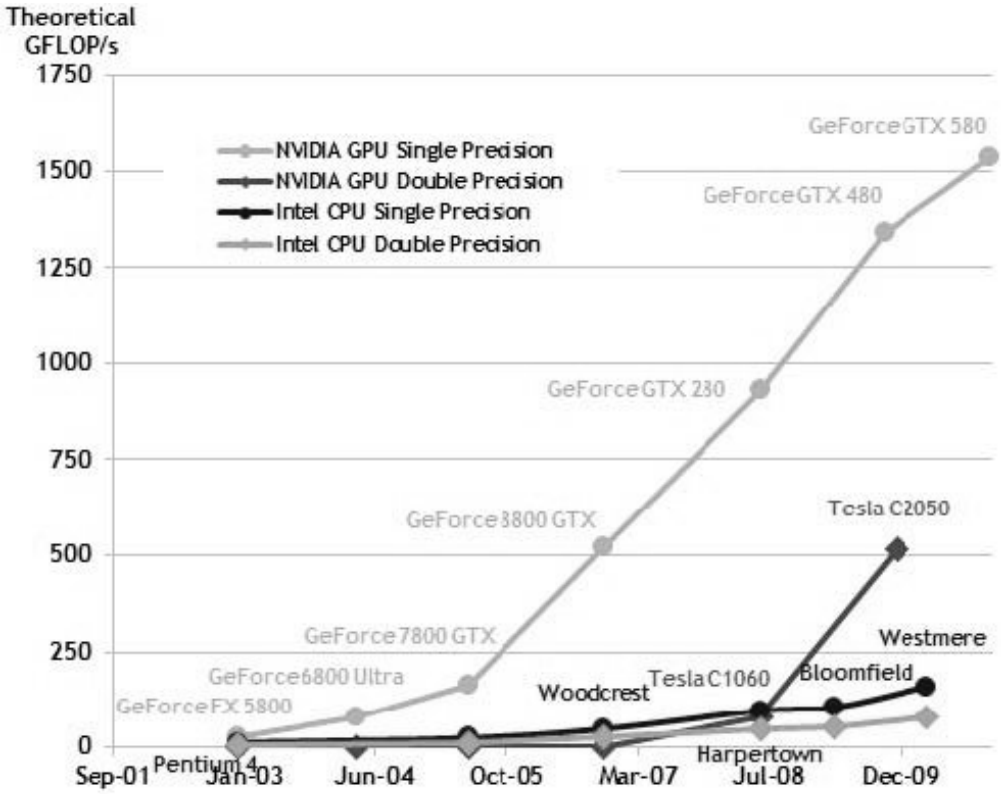

Figure 1. Theoretical FLOP rates of the GPU and CPU

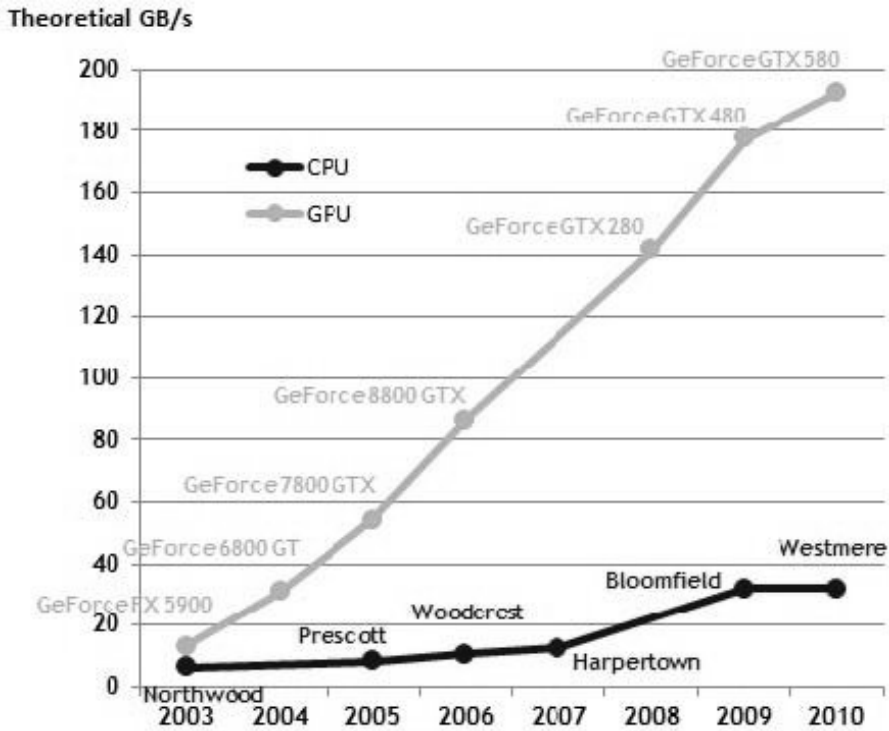

Figure 2. Theoretical memory bandwidth of the GPU

GPU can obtain such high FLOP rates because it is specialized for highly parallel computation and it has more transistors dedicated to data processing rather than flow control or data caching which is the case of a CPU. This is the main reason of the big difference in floating point computational power between GPU and CPU. Figure 3[11] (NVIDIA, 2011) shows this design shift in GPU.

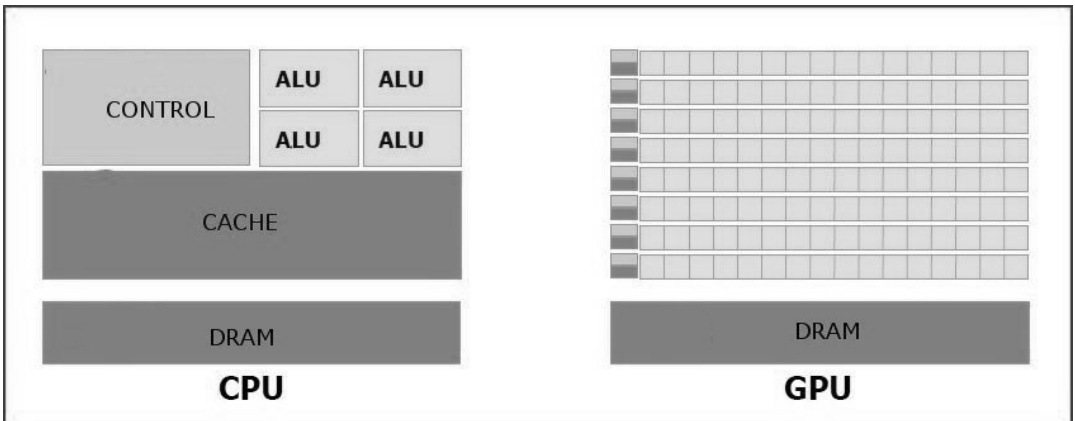

Figure 3. Design differences between GPU and CPU

---

[11] NVIDIA, (2011), NVIDIA CUDA C Programming Guide, version 4.0

GPUs are designed to solve problems that can be formulated as data-parallel computations – the same instructions are executed in parallel on many data elements with a high ratio between arithmetic operations and memory accesses. This is similar with the SIMD approach of the parallel computers taxonomy. Because the same instructions are executed on each of the data element there is no need for complicated flow control circuits and the memory latency can be hidden by arithmetic computations instead of using data cache.

Data parallel programming paradigm can be found in many applications like 3D rendering, image scaling, pattern recognition, video encoding and decoding, linear algebra routines, computational biology, computational finance or econometrics. All these applications can obtain very high speedups by mapping data elements to parallel processing threads that are executed in parallel by the GPU.

CUDA (Compute Unified Device Architecture) was introduced for the first time in 2006 by NVIDIA. It is a general purpose parallel programming architecture that uses the parallel compute engine in NVIDIA GPUs to solve complex computational problems in a more efficient way than a CPU does. At the time of its introduction CUDA supported only the C programming language, but nowadays it supports FORTRAN , C++ , Java, Phyton, etc.

The CUDA parallel programming model has three main key abstractions – a hierarchy of thread groups, shared memories, and barrier synchronization. These abstractions are exposed to the programmer as language extensions. They provide fine grain data parallelism and thread parallelism together with task parallelism that can be considered coarse grain parallelism.

The CUDA parallel programming model requires programmers to partition the problem to be solved into coarse tasks that can be independently executed in parallel by blocks of threads and each task is further divided into finer pieces of code that can be executed cooperatively in parallel by the threads within the block. This model allows threads to cooperate when solving each task, and also enables automatic scalability. Each block of threads can be scheduled for execution on any of the available processor cores, concurrently or sequentially. This allows a CUDA program to be executed on any number of processor cores. Figure 3[12] shows how a program is partitioned into blocks of threads each block being executed independently from each other. A GPU with more cores will execute the program in less time than a GPU with fewer cores.

---

[12] NVIDIA, (2011), NVIDIA CUDA C Programming Guide, version 4.0

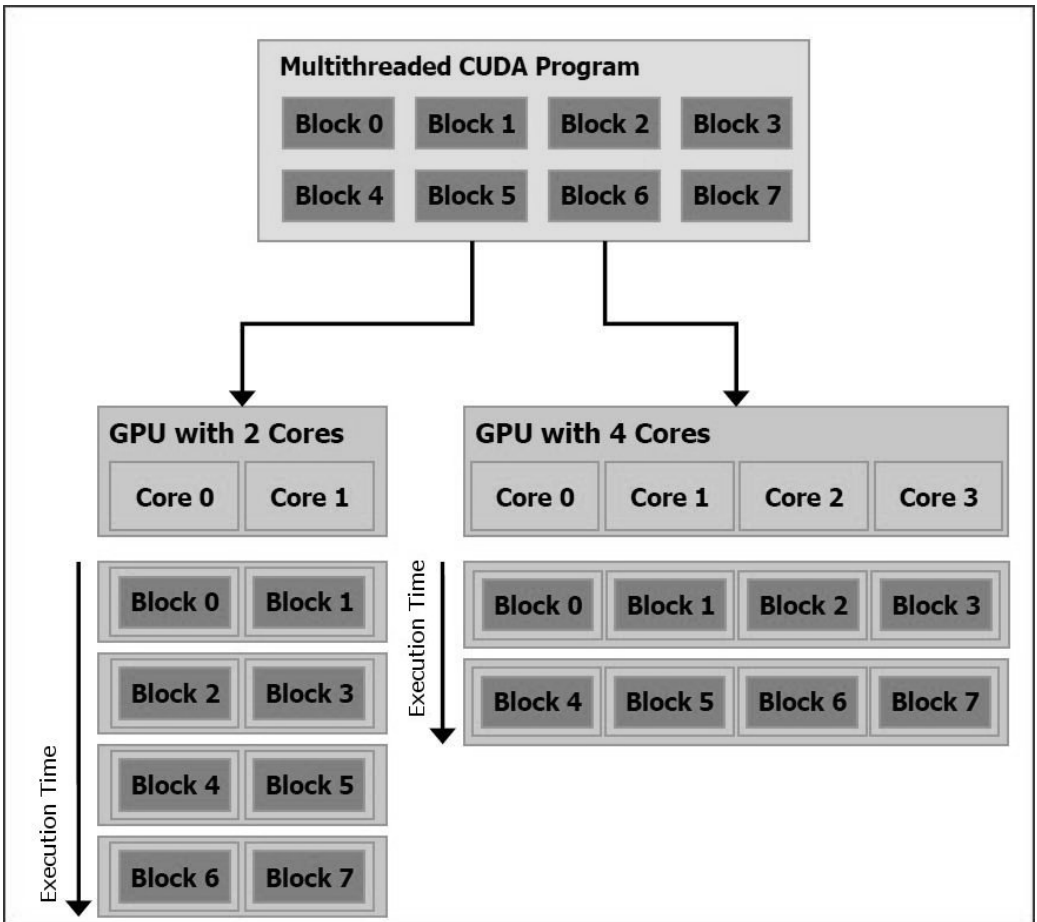

Figure 3. A multithreaded program divided into blocks that are allocated on 2 or 4 cores

The main C language extension of the CUDA programming model allows the programmer to define C functions, called **kernels**, that are executed N times in parallel by N different CUDA threads.

A CUDA **kernel** definition specifies the number of CUDA threads that execute that kernel for a given call. A unique *thread ID* identifies each thread that executes the kernel. This ID is accessible within the kernel through the built-in **threadIdx** variable. **ThreadIdx** variable is a 3-component vector, so that each thread can be identified using a one-dimensional, two-dimensional, or three-dimensional *index*. By indexing the threads in this way one can execute computations on data elements organized in a vector, matrix or 3D space.

The thread number per block is limited because all the threads of a block will be executed by one processor core and must share the limited memory resources of that processor core. This limitation on current GPUs implies that a thread block can contain up to 1024 threads. The blocks are structured into one-dimensional, two-dimensional, or three-dimensional grid of thread blocks. The number of thread blocks in a grid is limited by the size of the data being processed or the number of processors in the GPU.

In the CUDA programming model, threads can access data from multiple memory spaces during their execution life time. Each thread has its own private local memory, each thread block shares a memory visible to all the threads in that block and all threads have access to the same global memory.

CUDA threads are executed on a physically separate *device* that operates like a coprocessor to the *host* processor running the C program. The device is located on the GPU while the host is the CPU. The CUDA programming paradigm assumes that the kernels execute on GPU and the rest of the program executes on the CPU. It also assumes that that both the host and the device maintain their own separate memory spaces in DRAM, referred to as *host memory* and *device memory*. Using CUDA programming model a matrix multiplication code $C = A \times B$ can be structured like in the following example:

```
/* allocate memory for the matrices A, B, C in the host memory, A,B,C, being N x N matrices*/
host_matrix_A = (float*)malloc(N * N * sizeof(host_matrix_A[0]));
host_matrix_B = (float*)malloc(N * N * sizeof(host_matrix_B[0]));
host_matrix_C = (float*)malloc(N * N * sizeof(host_matrix_C[0]));
/* generate random test data */
randomTestData(host_matrix_A, host_matrix_B, host_matrix_C)
/* allocate memory for the matrices in the device memory space*/
cudaAlloc(N*N, sizeof(device_matrix_A[0]), (void**)&device_matrix_A);
cudaAlloc(N*N, sizeof(device_matrix_B[0]), (void**)&device_matrix_B);
cudaAlloc(N*N, sizeof(device_matrix_C[0]), (void**)&device_matrix_C);
/* copy the values from the host matrices to the device matrices */
cudaSetVector(N*N, sizeof(host_matrix_A[0]), host_matrix_A, 1, device_matrix_A, 1);
cudaSetVector(N*N, sizeof(host_matrix_B[0]), host_matrix_B, 1, device_matrix_B, 1);
cudaSetVector(N*N, sizeof(host_matrix_C[0]), host_matrix_C, 1, device_matrix_C, 1);
/* call the matrix multiplication routine*/
cudaMatMul(device_matrix_A, N, device_matrix_B, N, beta, device_matrix_C, N);
/* Copy the result from the device memory back to the host memory */
cudaGetVector(N*N, sizeof(host_matrix_C[0]), device_matrix_C, 1, host_matrix_C, 1)
```

**CUDA GPU applications**

Since its introduction in early 2007, a variety of applications have benefitted by the tremendous computational power of current GPUs. These benefits include few orders-of-magnitude speedups over the previous state-of-the-art implementations. Few of CUDA enabled applications are presented in the following.

Medical imaging is one of the earliest areas that benefited most from the CUDA programming model on GPU. Very large images from Ultrasound imaging devices or CT can be processed in only few minutes. For example[13] reports an improved algorithm for medical imaging reconstruction that benefits from the computational power of a Quadro FX5600 device that can shorten the processing time of an image, making the reconstruction practical for many clinical applications.

Another important challenge in the medical imaging field is the amount of data that is collected for a single patient. For a 4D (3D + time) computed tomography (CT) dataset can be of the resolution 512 x 512 x 512 x 20 and require more than 10 GB of memory storage. A first step in processing such an image is to apply image denoising. For a dataset of about 10-15 GB this can take several hours on the CPU, compared to 10-15 minutes on the GPU.

---

[13] Sam Stone, Justin P. Haldar, Stephanie C. Tsao, Wen-mei W. Hwu, Zhi-Pei Liang, Bradley P. Sutton, (2008), *Accelerating Advanced MRI Reconstructions on GPUs,* Proceedings of the 5th International Conference on Computing Frontiers, May 5-7, http://doi.acm.org/10.1145/1366230.1366274.

Computational fluid dynamics is another area that benefitted from the GPU developments. Several ongoing projects on Navier-Stokes models or Lattice Boltzmann methods have shown very large speedups using CUDA-enabled GPUs.

Other GPGPU applications that benefits from the advantages of CUDA programming model are:

- Linear algebra and large scale numerical simulations;
- Molecular dynamics, protein folding;
- Finance modeling;
- Signal processing (FFT);
- Raytracing;
- Physics simulation (fluid, cloth, collision);
- Speech and Image recognition;
- Databases;
- Sorting and searching algorithms;
- Astrophysics;
- Lattice QCD, theoretical physics.

The authors of this paper developed a CUDA based library that implements iterative algorithms for linear systems. Our library implements Jacobi, Gauss-Seidel, CG, GMRES and BiCGSTAB methods[14][15].

The general flow of a solver implemented in our library is:
Allocate memory for matrices and vectors in the host memory;
Initialize matrices and vectors in the host memory;
Allocate memory for matrices and vectors in the device memory;
Copy matrices from host memory to device memory;
Define the device grid layout:

○ Number of blocks

○ Threads per block

Execute the kernel on the device;
Copy back the results from device memory to host memory;
Memory clean up.

We've tested our iterative solver for both single precision and double precision floating point numbers. We used a computer with Intel Core2 Quad Q6600 procesor running at 2.4 Ghz, 4 GB of RAM and a NVIDIA GeForce GTX 280 graphics processing unit (GPU) with 240 cores running at 1296 MHz, 1GB of video memory and 141.7 GB/sec memory bandwith. We compared the results obtained using the CUDA code with a single threaded C implementation run on CPU that used the optimized ATLAS library (Whaley, 2001) as a BLAS implementation. Our performance tests show speedups of approximately 80 times for single precision floating point numbers and 40 times for double precision. These results show the immense potential of the GPGPU.

**Other GPGPU frameworks**

AMD Accelerated Parallel Processing (former ATI Stream technology) is a set of advanced hardware and software technologies that enable AMD graphics processors (GPU), working in together with the central processor (CPU) to accelerate applications (AMD, 2011). The APP programming model resembles the CUDA paradigm. It supports data-parallel and task-parallel programming models.

---

[14] Golub, G. H., and C. F. Van Loan, Matrix Computations (1996), Johns Hopkins Series in Mathematical Sciences, The Johns Hopkins University Press

[15] Saad, Y. (1996), Iterative Methods for Sparse Linear Systems, PWS Publishing Company.

OpenCL (Open Computing Language)[16] is the first open, standard for general-purpose parallel programming of heterogeneous systems. It tries to provide a unique programming environment for software developers to write portable code for servers, laptops, desktop computer systems and handheld devices using a both multi-core CPUs and GPUs.

OpenCL programs are divided in two parts: one that executes on the **device** (the GPU) and other that executes on the **host** (the CPU). The *device program* is the part of the code that uses GPU for parallel execution. Programmers have to write special functions called kernels which uses OpenCL Programming Language (an extension to the C programming language). These kerneles are scheduled to be executed on GPU. The host program offers an API so that the programmer can manage the execution of kernels on device. The host program can be programmed in C or C++ and it controls the OpenCL environment.

Open CL uses a SIMT (SINGLE INSTRUCTION MULTIPLE THREAD) model of execution that reflects how instructions are executed in the host. This means that the same code is executed in parallel by a different thread, and each **thread** executes the code with different data. Another concept of the Open CL is the work-item. The work-items are the equivalent of the CUDA threads being the basic execution entity. They have an ID accessible from the kernel. The ID is used to differentiate the data to processed by each work-item When a kernel is launched for execution a number of work-items specified by the programmer are also launched, each of them executing the same code. Work-items cooperate between them within a work-group. A work-group specifies the organization of the work-items which can be a N-dimensional gridwith N = 1, 2 or 3. Work-groups are equivalent to CUDA thread blocks. Work-groups also have an unique ID that can be referred from the kernel. The next level of organization of the device code is the ND-Range. It specify how work-groups are organized: as N-dimensional grid of work-groups, N = 1, 2 or 3. Table 1 summarizes the different but equivalent terms used by CUDa and OpenCL

Table 1. C for CUDA terminology versus OpenCL terminology

| CUDA         | OpenCL     |
|--------------|------------|
| Thread       | Work-item  |
| Thread block | Work-group |

**Conclusions**

This paper presents a new trend in parallel processing – GPGPU. This means using the GPU for execution of numerical intensive parts of general applications. GPUs have evolved over the years and now they are truly general processors and it outperforms CPU for numerical computations.

There are now two main frameworks that use the GPU for general applications: CUDA developed by NVIDIA and APP developed by AMD.

In the future the GPU devices will become more and more capable. This performance improvement will be mainly due to increased levels of parallelism. There is one limitation to the increase in the computational power of the GPU: the memory bandwidth that will play an increasing role.

Maybe one of the important developments in GPGPU field will be in the programmability of the GPUs. A particular attention should be paid to OpenCL framework which is a crossplatform and device-independent approach. This is very attractive to developers because is independent from hardware vendor.

---

[16] Khronos OpenCL Working Group (2009), The OpenCL Specification - Version 1.0. The Khronos Group, Tech. Rep.

**References**

- AMD (2008), ATI Stream Computing - Technical Overview. AMD, Tech. Rep.
- AMD (2011), *AMD Accelerated Parallel Processing – Programming Guide*.
- Barron, Iann M. (1978), D. Aspinall. ed. "The Transputer". The Microprocessor and its Application: an Advanced Course, Cambridge University Press.
- K. E. Batcher, (1980), Design of a Massively Parallel Processor, *IEEE Transactions on Computers*, Vol. C29, September, pp. 836–840.
- Golub, G. H., and C. F. Van Loan, Matrix Computations (1996), Johns Hopkins Series in Mathematical Sciences, The Johns Hopkins University Press.
- Allan Gottlieb, Ralph Grishman, Clyde P. Kruskal, Kevin P. McAuliffe, Larry Rudolph, Marc Snir, (1982), *The NYU Ultracomputer—designing a MIMD, shared-memory parallel machine*, ISCA '82 Proceedings of the 9th annual symposium on Computer Architecture, pp. 27 – 42.
- Harris, Mark J., William V. Baxter III, Thorsten Scheuermann, and Anselmo Lastra.( 2003), Simulation of Cloud Dynamics on Graphics Hardware. In *Proceedings of the IGGRAPH/Eurographics Workshop on Graphics Hardware* 2003, pp. 92-101.
- INTEL (2012), Microprocessor Quick reference guide.
- Khronos OpenCL Working Group (2009), The OpenCL Specification - Version 1.0. The Khronos Group, Tech. Rep.
- NVIDIA, (2010), CUDA C Programming Guide, Version 4.0.
- NVIDIA (2007) CUDA – CUBLAS Library.
- NVIDIA, (2011), NVIDIA CUDA C Programming Guide, version 4.0.
- Oancea, Bogdan, Ion. Gh. Rosca, Tudorel Andrei, Andreea Iluzia Iacob,(2011), Evaluating Java performance for linear algebra numerical computations, Procedia Computer Science, vol. 3, pp. 474-478.
- Saad, Y. (1996), Iterative Methods for Sparse Linear Systems, PWS Publishing Company.
- Sam Stone, Justin P. Haldar, Stephanie C. Tsao, Wen-mei W. Hwu, Zhi-Pei Liang, Bradley P. Sutton, (2008), *Accelerating Advanced MRI Reconstructions on GPUs,* Proceedings of the 5th International Conference on Computing Frontiers, May 5-7, http://doi.acm.org/10.1145/1366230.1366274.
- Lewis W. Tucker, George G. Robertson, (1988), Architecture and Applications of the Connection Machine, *Computer*, vol. 21, no. 8, pp. 26–38.
- Whaley, R. C., A. Petitet, and J. Dongarra (2001), "Automated Empirical Optimization of Software and the ATLAS project", Parallel Computing, 27(1-2), 3-35.